\documentclass[11pt,twoside]{article}

\usepackage{graphics}
\usepackage{graphicx}
\usepackage{asp2006}
\usepackage{epsf}
\usepackage{psfig}
\usepackage{lscape}
\begin{document}
\title{The Spectral Energy Distributions of AGN with Double-Peaked Balmer Lines}   
\author{I. Strateva\altaffilmark{1}, W. N. Brandt\altaffilmark{2}, M. Eracleous\altaffilmark{2,3}, G. Garmire\altaffilmark{2}, S. Komossa\altaffilmark{1}}   
\altaffiltext{1}{MPI f\"ur extraterrestrische Physik, Postfach 1312, 85741 Garching, Germany}  
\altaffiltext{2}{Dept. of Astronomy and Astrophysics, 525 Davey Lab., University Park, PA 16802, USA}  
\altaffiltext{3}{Center for Gravitational Wave Physics, 104 Davey Lab., University Park, PA 16802, USA}

\begin{abstract} 
  We summarize the optical, UV, and X-ray properties of double-peaked
  emitters -- AGN with double-peaked Balmer emission lines believed to
  originate in the AGN accretion disk. We focus on the X-ray
  spectroscopic results obtained from a new sample of the 16 broadest
  Balmer line AGN observed with \emph{Chandra} and \emph{Swift}.
\end{abstract}



\section{Introduction}
Forty years of research on the Broad Line Regions (BLRs) of Active
Galactic Nuclei (AGN) has not yielded a clear picture of the
structures and detailed physical conditions involved in setting the
observed line profiles and strengths. Our current understanding
involves contributions from an outflowing, likely clumpy,
magnetohydrodynamically or line-driven wind, and an
accretion disk \citep[e.g,][]{C06}. The BLR is photoionized and must intercept
at least 10\% of the radiation from the central engine to produce
lines of the observed strengths. If the accretion disk contributes to
the BLR emission from AGN, we expect to see a clear signature of this
contribution in the form of a double-peaked line profile, distorted by
the effects of Doppler boosting and gravitational redshift.

\section{Optical and UV Properties of Double-Peaked Emitters}   
About 3\% of optically selected AGN have H$\alpha$ lines with the
characteristic double-peaked shape consistent with accretion-disk
emission \citep[][hereafter S03]{S03}.  Among broad-line radio
galaxies this fraction is up to 20\% \citep[][hereafter E03]{E03},
suggesting that Balmer-line accretion-disk emission is more prevalent
among AGN with jets and lower luminosities/accretion rates.

As a class double-peaked emitters are characterized by broader Balmer
lines (up to 40,000\,km\,s$^{-1}$ full width at half maximum [FWHM]),
similar UV/optical luminosities to those of other AGN in the same
redshift range, and a higher incidence of LINER characteristics -- for
example, larger low-ionization line ratios and low-ionization line
equivalent widths ([O\,I]\,6300\AA/[O\,III]\,5007\AA, [S\,II]\,EWs,
etc.; E03, S03).  \citet{L06} report the black-hole masses of 8
double-peaked emitters based on the stellar velocity dispersions of
their host-galaxy bulges and find that the accretion rates as a
function of Eddington vary considerably even in this small sample,
with $10^{-5}<L/L_{\textnormal{\scriptsize{Edd}}}<0.1$.  UV
observations of double-peaked emitters often show single-peaked
high-ionization lines (e.g. C\,III], CIV, Ly$\alpha$;
\citeauthor[e.g.,][]{H96}\,1996), and are likely dominated by emission from the
outflowing BLR component.
\begin{figure}[!h]
\includegraphics[scale=0.62]{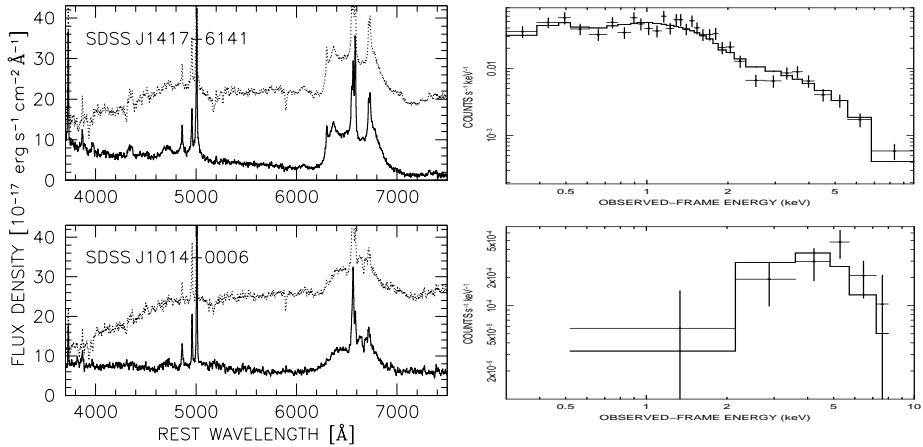}
\caption{Optical (\emph{left panels;} both the AGN and AGN+host-galaxy
  spectra are shown) and X-ray spectra (\emph{right panels}) of two of
  the \emph{Swift} targets.}
\end{figure}

Axisymmetric accretion-disk fits to the H$\alpha$ line profiles of
about 3 dozen objects from the E03 and S03 samples constrain the
accretion-disk line emission to originate between a few hundred and a
few thousand $R_G$ (gravitational radii) from the center. The outer
radii of emission, typically $\la 4000$\,$R_G$\footnote{The outer
  emission radii are often poorly constrained due to a degeneracy
  between the disk outer radius and the power-law slope of the
  illuminating radiation.}, are consistent with the onset of disk self
gravity. The disk inclinations range between \hbox{$20\deg<i<75\deg$}
(from the accretion-disk axis), with the inclination distribution
suggestive of the presence of coplanar obscuration, whose probability
becomes exponentially higher for large inclinations. This is consistent
with clumpy unification models (see the Elitzur, Sturm, and Hao
contributions in this volume).

\section{X-ray Properties of Double-Peaked Emitters}   
\citet{S06} [S06] studied the X-ray emission of 39 double-peaked
emitters, serendipitously detected in \emph{ROSAT} (22/39),
\emph{Chandra} and \emph{XMM-Newton} observations. As a class the
double-peaked emitters showed similar UV/optical-to-X-ray ratios
($\alpha_{\textnormal{\scriptsize{ox}}} = 0.3838 \log
[F_{\textnormal{\scriptsize{2\,keV}}}/{F_{\textnormal{\scriptsize{2500\,\AA}}}}]$)
and X-ray power-law spectral indices to other broad-line AGN. In
practically all cases (see Fig.~3b) a comparison of the H$\alpha$
emission strength with the viscous power available locally from the
accretion disk supports the need
for external illumination of the disk to produce lines of the observed
strength.
\section{New \emph{Chandra} and \emph{Swift} Observations}   
Our \emph{Chandra}(5/16)/\emph{Swift}(11/16) sample consists of the 16
double-peaked emitters with the broadest Balmer lines
(FWHM$_{\textnormal{\scriptsize{H$\alpha$}}}>15,000$\,km\,s$^{-1}$)
ever observed. The Balmer-line emission in these AGN originates from
the immediate vicinity of the black hole, immediately adjacent to the
region responsible for the X-ray emission and the external
illumination.  Fig.~1 shows the optical and X-ray spectra of two
radio-quiet [RQ] double-peaked emitters. Despite their comparable
4200\AA\ luminosities their X-ray spectra are strikingly different --
with large cold absorption modifying the power-law emission in one
case, but not in the other ($5\times10^{20}$\,cm$^{-2}$ and
$10^{23}$\,cm$^{-2}$ for the top and bottom spectra, respectively).
Half of the \emph{Chandra}/\emph{Swift} sample objects show intrinsic
X-ray absorption.
\begin{figure}[!ht]
\plottwo{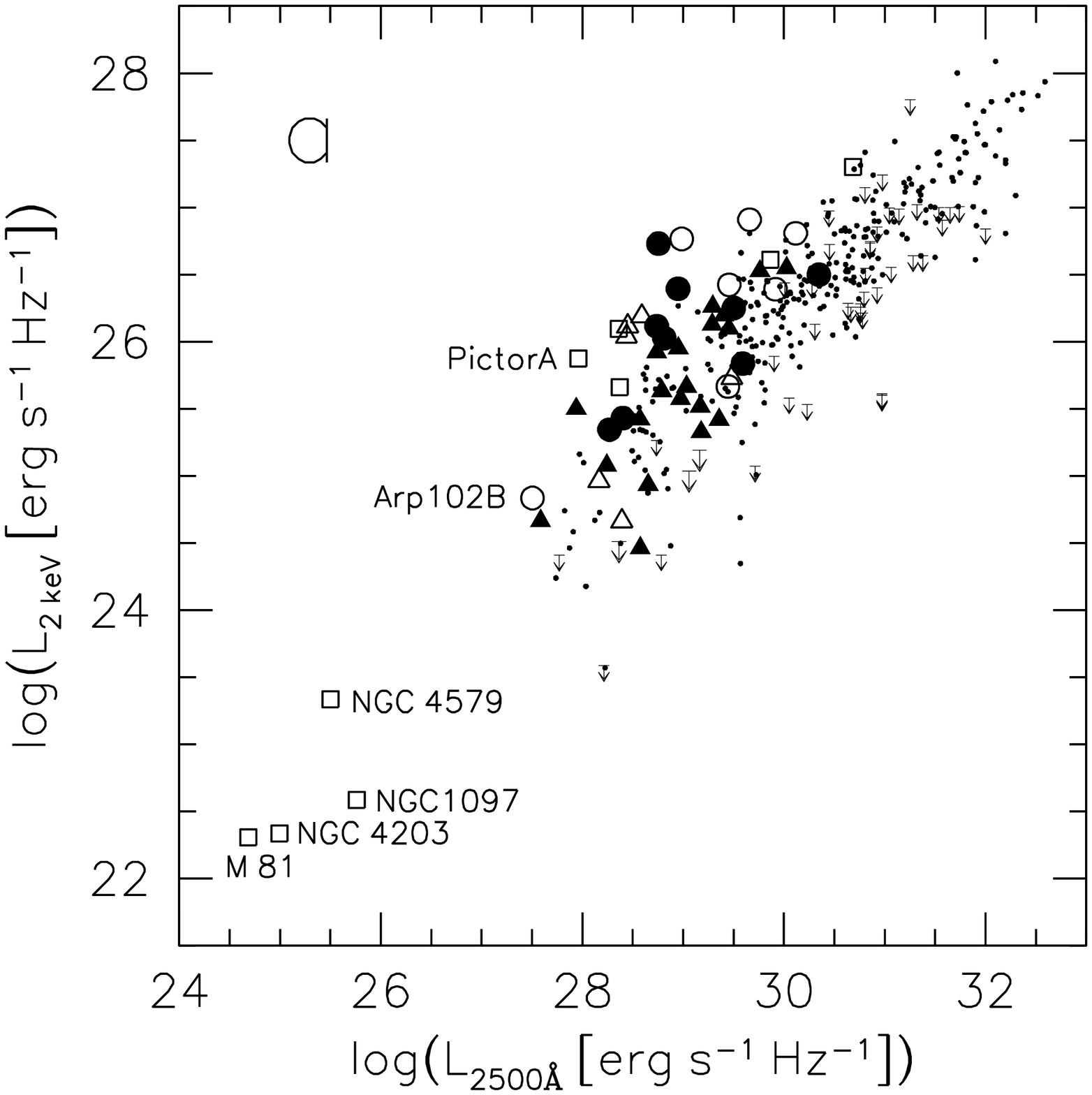}{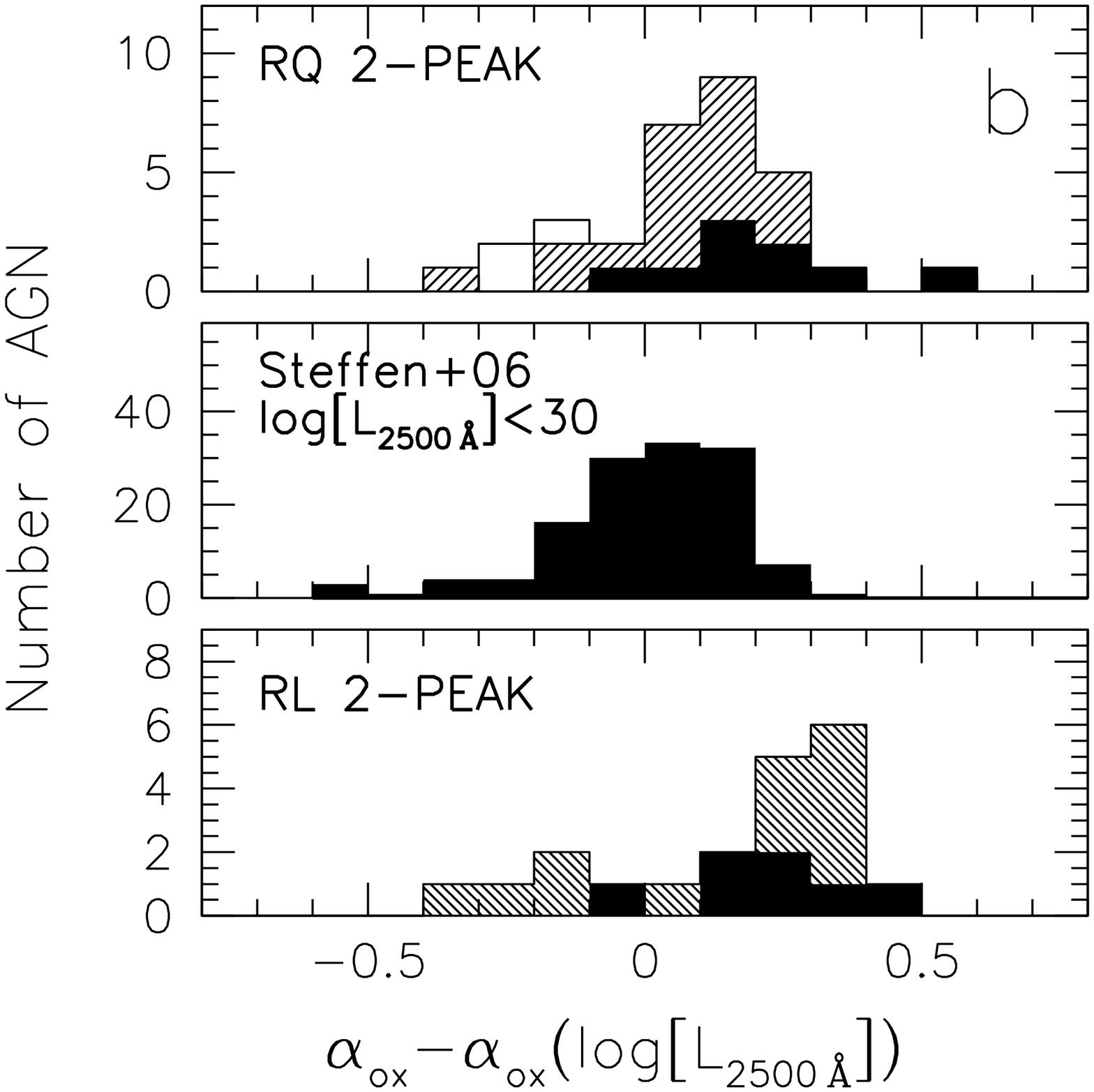}
\caption{\emph{Left:} The 2500\AA\ vs. 2\,keV monochromatic
  luminosities of the \emph{Chandra}/\emph{Swift} targets (circles),
  the double-peaked emitters from S06 (triangles and squares), and the
  general SDSS AGN population (small dots). The open symbols denote RL
  AGN, the solid symbols RQ AGN, and the arrows upper limits.
  \emph{Right:} The $\alpha_{\textnormal{\scriptsize{ox}}}$ residual
  distributions for the RQ double-peaked emitters (top), the general
  SDSS lower-luminosity population (middle) and the RL double-peaked
  emitters (bottom). The solid histograms in the top and bottom panels
  show only the \emph{Chandra}/\emph{Swift} samples, while the open
  histogram indicates X-ray limits.}
\end{figure}
\begin{figure}[!ht]
\plottwo{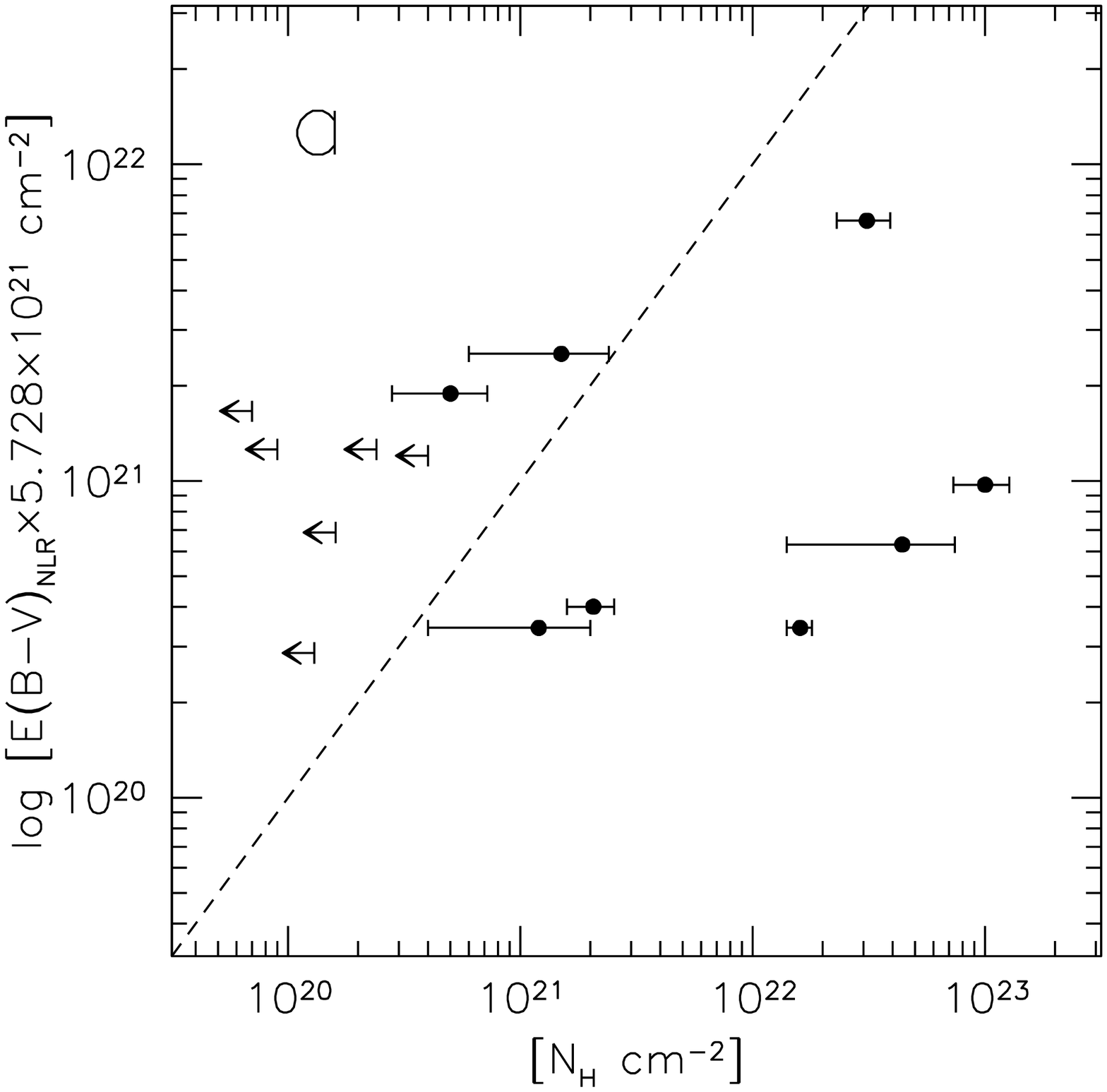}{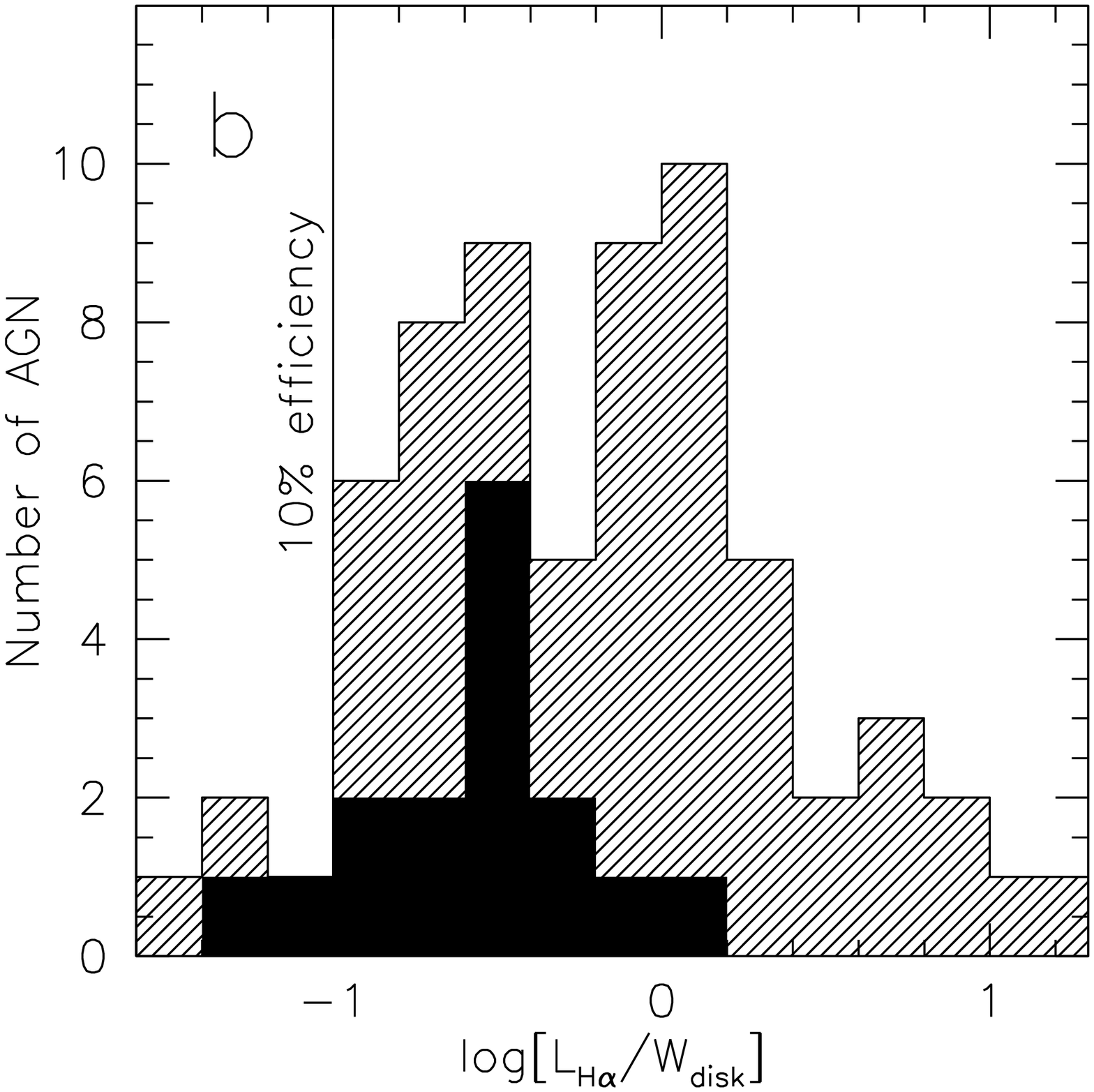}
\caption{\emph{Left:} The intrinsic X-ray absorption (arrows indicate
  limits based on the Galactic absorption) vs. the Hydrogen column
  inferred from the NLR Balmer decrement; the dashed line is a line of equality.
  \emph{Right:} The distribution of the ratio of line-strength
  ($L_{\textnormal{\scriptsize{H$\alpha$}}}$) and locally available
  viscous power ($W_{\textnormal{\scriptsize{disk}}}$) for the
  \emph{Chandra}/\emph{Swift} targets (solid histogram) and the E03
  and S06 objects (hatched). A ratio larger than 10\% (solid line)
  indicates the need for external illumination.}
\end{figure}

The X-ray power-law spectral indices of extremely broad double-peaked
emitters are similar to those of other double-peaked emitters and
normal AGN with comparable radio-loudness. Unlike the general sample
of double-peaked emitters, however, the broadest Balmer-line
double-peaked emitters appear \hbox{X-ray} brighter relative to other AGN
with comparable UV emission (see Fig.~2). In Fig.~2b we quantify this
difference by comparing the $\alpha_{\textnormal{\scriptsize{ox}}}$
residual distributions\footnote{The
  $\alpha_{\textnormal{\scriptsize{ox}}}$ residuals are obtained by
  subtracting the $\alpha_{\textnormal{\scriptsize{ox}}}$ dependence
  on luminosity measured by \citet{St06} for RQ AGN,
  $\alpha_{\textnormal{\scriptsize{ox}}}=-0.137\log[L_{\textnormal{\scriptsize{2500\,\AA}}}]+2.638$,
  from the measured $\alpha_{\textnormal{\scriptsize{ox}}}$.} for RQ
and radio-loud [RL] double-peaked emitters to those of normal SDSS AGN
with comparable luminosities. RL AGN are brighter in the \hbox{X-rays} than
similar UV luminosity RQ AGN regardless of their Balmer-line shapes
(i.e. they have larger $\alpha_{\textnormal{\scriptsize{ox}}}$
residuals). After the addition of the broadest Balmer-line
double-peaked emitters to the objects in S06, we now find a
statistically significant difference (99\%) for the RQ double-peaked
emitters, which are also X-ray brighter.

Fig.~3a shows the X-ray absorbing-gas column vs. the hydrogen column
estimated based on the optical reddening (via the narrow line region
[NLR] Balmer-line decrements).  In a handful of double-peaked emitters
(those close to the dashed line in Fig.~3a) the observations are
consistent with a single absorber with gas-to-dust ratio similar to
that of the interstellar medium in our Galaxy (e.g. the AGN host galaxy). In at least
3 cases the X-ray absorbing column is much higher, possibly indicating
a dust-free or a partial absorber on scales much smaller than the NLR.



\end{document}